\documentclass[twocolumn,preprintnumbers,superscriptaddress,amsmath,amssymb,prb,noeprint]{revtex4-1}

\usepackage{braket}
\usepackage{miller}
\usepackage{changes}

\usepackage[textsize=tiny]{todonotes}
\usepackage{graphicx}

\usepackage[utf8]{inputenc}
\usepackage{multirow}
\usepackage{float}

\usepackage{bm}
\usepackage{verbatim}\label{key}
\usepackage{xfrac}
\usepackage{array}
\usepackage{siunitx}
\usepackage{ulem}
\usepackage[colorlinks,allcolors=blue]{hyperref}

\begin{document}
	
	\title{Measuring topological invariants in polaritonic graphene}

	\author{P. St-Jean}
	\affiliation{Centre de Nanosciences et de Nanotechnologies (C2N), CNRS - Universit\'e Paris-Sud / Paris-Saclay, Palaiseau, France}
	
	\author{A. Dauphin}
	\affiliation{ICFO-Institut de Ciencies Fotoniques, The Barcelona Institute of Science and Technology, Barcelona, Spain}
	
	\author{P. Massignan}
	\affiliation{ICFO-Institut de Ciencies Fotoniques, The Barcelona Institute of Science and Technology, Barcelona, Spain}
	\affiliation{Departament de F\'isica, Universitat Polit\`ecnica de Catalunya, Barcelona, Spain}
	
	\author{B. Real}
	\affiliation{Physique des Lasers, Atomes et Mol\'ecules (PhLAM), CNRS - Universit\'e de Lille, Lille, France}
	
	\author{O. Jamadi}
	\affiliation{Physique des Lasers, Atomes et Mol\'ecules (PhLAM), CNRS - Universit\'e de Lille, Lille, France}
	
	\author{M. Milicevic}
	\affiliation{Centre de Nanosciences et de Nanotechnologies (C2N), CNRS - Universit\'e Paris-Sud / Paris-Saclay, Palaiseau, France}
	
	\author{A. Lema\^itre}
	\affiliation{Centre de Nanosciences et de Nanotechnologies (C2N), CNRS - Universit\'e Paris-Sud / Paris-Saclay, Palaiseau, France}
	
	\author{A. Harouri}
	\affiliation{Centre de Nanosciences et de Nanotechnologies (C2N), CNRS - Universit\'e Paris-Sud / Paris-Saclay, Palaiseau, France}
	
	\author{L. Le Gratiet}
	\affiliation{Centre de Nanosciences et de Nanotechnologies (C2N), CNRS - Universit\'e Paris-Sud / Paris-Saclay, Palaiseau, France}
	
	\author{I. Sagnes}
	\affiliation{Centre de Nanosciences et de Nanotechnologies (C2N), CNRS - Universit\'e Paris-Sud / Paris-Saclay, Palaiseau, France}
	
	\author{S. Ravets}
	\affiliation{Centre de Nanosciences et de Nanotechnologies (C2N), CNRS - Universit\'e Paris-Sud / Paris-Saclay, Palaiseau, France}
	
	\author{J. Bloch}
	\affiliation{Centre de Nanosciences et de Nanotechnologies (C2N), CNRS - Universit\'e Paris-Sud / Paris-Saclay, Palaiseau, France}
	
	\author{A. Amo}
	\affiliation{Physique des Lasers, Atomes et Mol\'ecules (PhLAM), CNRS - Universit\'e de Lille, Lille, France}
	
\begin{abstract}
	Topological materials rely on engineering global properties of their bulk energy bands called topological invariants. These invariants, usually defined over the entire Brillouin zone, are related to the existence of protected edge states. However, for an important class of Hamiltonians corresponding to 2D lattices with time-reversal and chiral symmetry (e.g. graphene), the existence of edge states is linked to invariants that are not defined over the full 2D Brillouin zone, but on reduced 1D sub-spaces. Here, we demonstrate a novel scheme based on a combined real- and momentum-space measurement to directly access these 1D topological invariants in lattices of semiconductor microcavities confining exciton-polaritons. We extract these invariants in arrays emulating the physics of regular and critically compressed graphene sucht that Dirac cones have merged. Our scheme provides a direct evidence of the bulk-edge correspondence in these systems, and opens the door to the exploration of more complex topological effects, for example involving disorder and interactions.
\end{abstract}
	
	\maketitle
	
	\setcounter{topnumber}{2}
	\setcounter{bottomnumber}{2}
	\setcounter{totalnumber}{4}
	\renewcommand{\topfraction}{0.85}
	\renewcommand{\bottomfraction}{0.85}
	\renewcommand{\textfraction}{0.15}
	\renewcommand{\floatpagefraction}{0.7}	
	
	
	
	Topological bands are characterized by integer-valued quantities, called topological invariants, that are typically defined as the integral of a local property (e.g. the Berry curvature) over the full Brillouin zone (BZ). The hallmark of these invariants is their robustness against local perturbations, which endows topological matter with properties that are insensitive to certain types of disorder and defects\cite{Hasan2010, Qi2011}. One notable example is provided by the edge conductivity plateaus in the quantum Hall effect that can be linked to a topological invariant called the Chern number\cite{Thouless1982}.
	
	A distinct situation arises in 2D crystals presenting time-reversal and chiral (or sub-lattice) symmetry, such as honeycomb, Lieb and Kagome lattices. The bands of these materials are either ungapped or present a globally vanishing Berry curvature; they thus can't be described by a non-zero first-order topological invariant defined over the entire BZ such as the Chern number. Yet, for well-defined crystalline terminations, these materials present edge states that can be linked to topological invariants defined over reduced (1D) sub-spaces of the BZ\cite{Ryu2002, Delplace2011}.
	
	So far, these 1D invariants have solely been determined indirectly by probing the emergence of edge states in honeycomb lattices \cite{Plotnik2014, Milicevic2015}. Yet, they are bulk properties, and one should be able to extract them without relying on measurements localized at an interface. This is critical in several situations where edges are difficult to probe, e.g. in disordered lattices\cite{Meier2018, Stutzer2018}.
	
	Extracting topological invariants from the bulk is a challenging task in solid-state crystals. Hence, artificial materials, e.g. arrays of cold atoms\cite{Goldman2016, Cooper2019} or photonic crystals\cite{Lu2014, Ozawa2019, Klembt2018}, are particularly appealing as they allow accessing topological properties of band structures through optical means\cite{Atala2013, Aidelsburger2015, Duca2015, Flaschner2016, Wimmer2017, Gianfrate2020}. However, techniques previously developed for extracting topological invariants were all designed for invariants defined in a parameter space of the same dimension as the underlying lattice.
	
	In this work, we propose and experimentally demonstrate a powerful technique that allows measuring topological invariants in 2D lattices with chiral symmetry. This technique, based on the concept of mean chiral displacement\cite{Cardano2017, Maffei2018}, consists in optically probing the spatial distribution of a wave-packet for a specific momentum component. The experimental implementation of this scheme is realized in patterned semiconductor microcavities confining exciton-polaritons\cite{Carusotto2013}. This system is particularly well suited for this purpose, as its dissipative nature allows accessing both momentum and real space profiles of Bloch modes with simple imaging techniques. Using polaritonic lattices emulating regular\cite{Jacqmin2014} and critically compressed\cite{Milicevic2019a} graphene (i.e. where Dirac cones have merged), we measure these 1D topological invariants and thus provide a direct evidence of the bulk-edge correspondence in these chiral systems. 
	
	Interestingly, our results show that topological invariants can be accessed in non-hermitian systems subject to moderate dissipation. Moreover, they pave the way to the exploration of more complex phenomena in 2D chiral lattices. Indeed, although the present work deals with polaritons in the linear regime thus essentially behaving as photons, their hybrid light-matter nature could allow probing, at higher excitation powers, the effect of interactions on topological invariants\cite{DeLeseleuc2019}.\\
	
	\begin{figure}
		\centering
		\includegraphics[trim=0cm 0cm 0cm 0.cm, width=0.9\columnwidth]{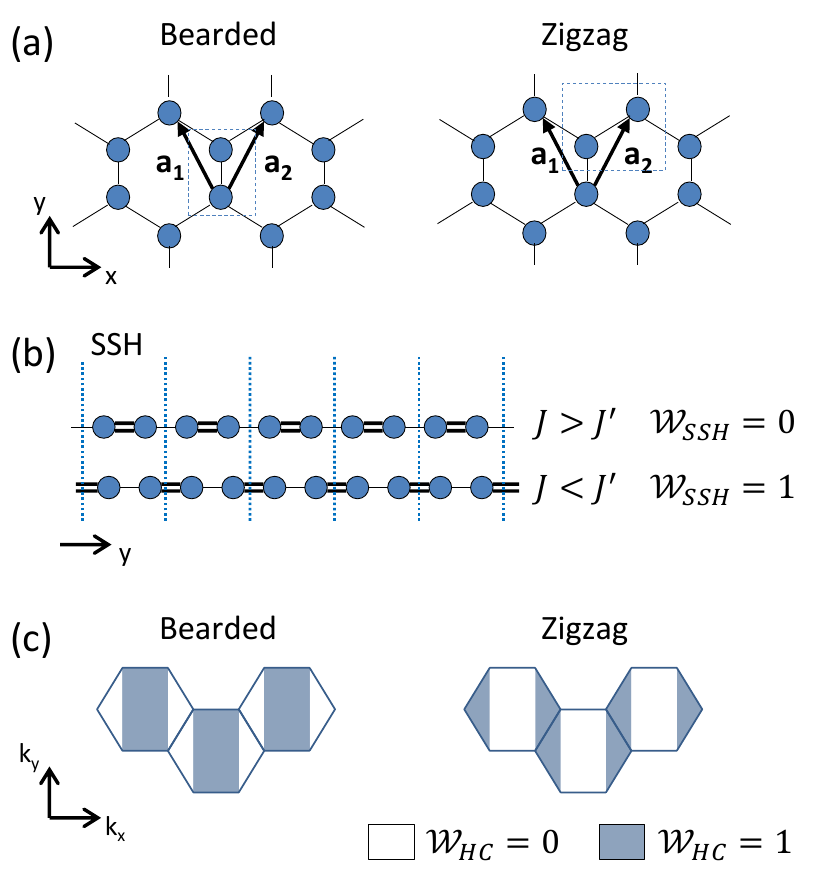}
		\caption{\textbf{Defining winding numbers in graphene.} (a) Definition of the unit cell compatible with bearded and zigzag terminations. $\mathrm{\textbf{a}_{1,2}}=\{ \pm 3 / \sqrt{2}, 3/2 \}$ are the basis vectors of the lattice. (b) Schematic representation of the two topological phases of the SSH model. The dotted lines indicate the boundaries of unit cells. (c) Evolution of the winding number as $k_{x}$ spans the BZ for bearded and zigzag terminations. Red and blue regions correspond respectively to a winding number of $\mathcal{W_{\mathrm{HC}}}=1$ and $\mathcal{W_{\mathrm{HC}}}=0$.}
		\label{sshGraphene}
	\end{figure}
	
	\noindent\textbf{1D topological invariants of graphene}
	
	Chiral symmetry describes periodic arrays formed of two sub-lattices with identical on-site energies, and couplings between distinct sub-lattices only\cite{Schnyder2008}. Honeycomb lattices, i.e. graphene-like materials, are one of the most notable examples of 2D lattices presenting this symmetry. For specific terminations, graphene presents edge states that can be linked to topological invariants defined in 1D sub-spaces of the BZ, corresponding to cuts along specific momentum directions\cite{Ryu2002, Delplace2011}. Before going into the details of the experimental extraction of these invariants, we first recall how they can be defined and how they can be related to the existence of edge states.
	
	\begin{figure*}
		\centering
		\includegraphics[trim=0cm 0cm 0cm 0cm,  width=0.9\textwidth]{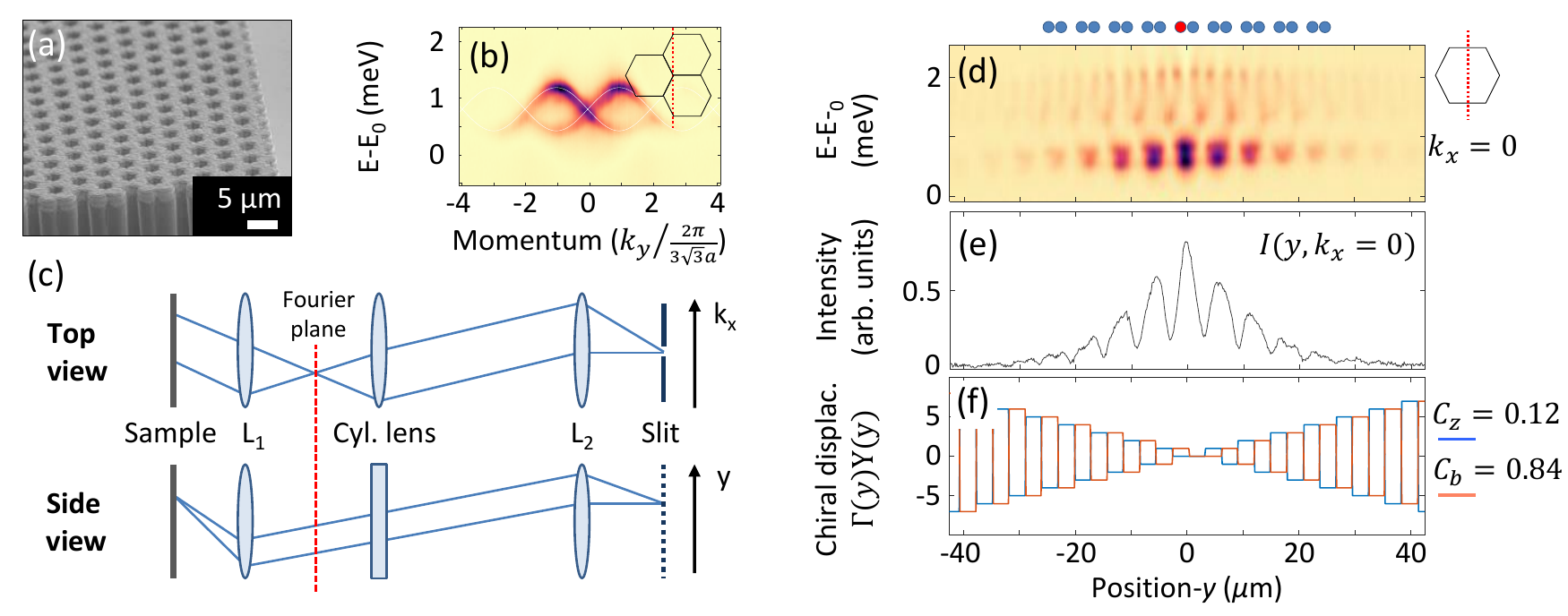}
		\caption{\textbf{Measuring the mean chiral displacement in graphene.} (a) Scanning electron microscopy (SEM) image of a honeycomb lattice of coupled micropillars. (b) Momentum-resolved emission spectra of a polaritonic graphene lattice. $E_{0}=1.571~\mathrm{eV}$. The cut in the BZ along which the image is taken is depicted in the inset. (c) Schematic representation of the setup, upper and lower panels depict top and side views. (d) Spatially resolved steady-state emission spectra (along $y$) for a value of $k_{x}\sim 0$. The position of the different sites of the effective SSH lattice is indicated above, where the pumped pillar is in red. (e) Emission intensity integrated over both bands as a function of spatial position. (f) Definition of the chiral displacement $\Gamma Y$ as a function of spatial position. The blue (red) curves correspond to definition of the unit cell compatible with zigzag (bearded) edges. The values of the mean chiral displacements $\mathcal{C}_{z}$ and $\mathcal{C}_{b}$ extracted from the spectrum presented in Panel (e) are given on the right of this panel.}
		\label{grapheneMCD}
	\end{figure*}
	
	In the sub-lattice basis, the tight-binding Hamiltonian in momentum space describing a particle hopping on a honeycomb lattice is given by:
	
	\begin{equation}
	H_{\mathrm{HC}}(\vec{k})=-j\left(
	\begin{array}{cc}
	0 & g(\vec{k})\\
	g(\vec{k})^{\dag} & 0
	\end{array}
	\right),
	\label{hamiltonianGraph}
	\end{equation}
	
	\noindent where $g(\vec{k})=1+e^{-i\vec{k}\cdot\vec{a_{1}}}+e^{-i\vec{k}\cdot\vec{a_{2}}}$, with $j$ the nearest-neighbour hopping amplitude. $H_{\mathrm{HC}}$ is defined for a unit cell compatible with bearded edges when considering a finite-sized ribbon along $y$, with periodic boundary conditions along $x$, see Fig.~\ref{sshGraphene} (a); for zigzag edges, $g(\vec{k})=1+e^{-i\vec{k}\cdot\vec{a_{2}}}(1+e^{i\vec{k}\cdot\vec{a_{1}}})$. For simplicity, we only consider bearded and zigzag terminations, but the argument and experimental technique developed in this work can be extended to arbitrary edges\cite{Delplace2011}.
	
	To better understand how 1D topological invariants can be defined for this Hamiltonian, it is insightful to separate explicitly the momentum components parallel ($k_{x}$) and perpendicular ($k_{y}$) to the edge in $g(\vec{k})$:
	
	\begin{equation}
	g(\vec{k})=J(k_{x})+J'(k_{x})e^{-i\tfrac{3}{2}ak_{y}}
	\end{equation}
	
	\noindent where $J(k_{x})=1$ and $J'(k_{x})=2\,\mathrm{cos}(\frac{\sqrt{3}ak_{x}}{2})$ for bearded edges, and vice-versa for zigzag edges (each term also includes a global phase factor $e^{-i\frac{\sqrt{3}ak_{x}}{2}}$). For each momentum component $k_{x}$, $H_{\mathrm{HC}}$ is thus isomorphic, along $y$, to the Hamiltonian of the well-known Su-Schreefer-Heeger model (SSH) which represents a 1D dimer chain with different intra-cell ($J$) and inter-cell ($J'$) coupling energies (see Fig.~\ref{sshGraphene} (b)). For this SSH Hamiltonian, it is possible to define a topological invariant called the winding number $\mathcal{W_{\mathrm{SSH}}}$ which is equal to 0 for the dimerization $J>J'$, and equal to 1 for $J<J'$. In the latter case, the non-zero topological invariant is linked to the existence of 0-energy edge states.

	In a similar manner, it is possible to define a winding number of $H_{\mathrm{HC}}$ for each value of $k_{x}$:
	
	\begin{equation}
	\mathcal{W_{\mathrm{HC}}}(k_{x})=\frac{1}{2\pi}\int_{BZ}\mathrm{d}k_{y}\frac{\partial\phi(\vec{k})}{\partial k_{y}},
	\end{equation}
	
	\noindent where $\phi(\vec{k})=\mathrm{arg}(g(\vec{k}))$. This 1D topological invariant is equal to the geometric (or Zak) phase picked up by a particle spanning the BZ along $k_{y}$, divided by $\pi$. It can only take two values: 0 for $|J'(k_{x})/J(k_{x})|<1$ and 1 for $|J'(k_{x})/J(k_{x})|>1$, which are equal to the number of edge states for the corresponding momentum $k_{x}$.
	
	Fig.~\ref{sshGraphene} (c) shows calculated values of $\mathcal{W_{\mathrm{HC}}}$ as a function of $k_{x}$ for honeycomb Hamiltonians where the unit cell is compatible with bearded and zigzag terminations. Blue (white) areas correspond to regions where $\mathcal{W_{\mathrm{HC}}}=1~(=0)$. For each termination, blue areas indeed correspond to regions of the BZ where the lattice exhibits 0-energy edge states\cite{Nakada1996, Delplace2011, Plotnik2014, Milicevic2015}. Transitions from one value of the winding number to the other occur at the positions of the Dirac cones: this reflects the fact that the gap needs to close and re-open in order to change a topological invariant.
	
	This mapping between honeycomb and SSH Hamiltonians has only been used so far as a theoretical tool, e.g. for understanding the emergence of edge states\cite{Plotnik2014, Milicevic2015, Ruffieux2016}. Our aim is to directly access these effective SSH Hamiltonians, and measure their winding number as a function of $k_{x}$.\\
	
	\noindent\textbf{Extracting winding numbers of graphene}
	
	It was shown in Refs. [\onlinecite{Maffei2018, Cardano2017}] that the winding number of any 1D chiral Hamiltonian can be extracted by probing the spatial evolution of an initially localized wave-packet, and computing a quantity called the mean chiral displacement. Hereafter, we demonstrate how this concept can be harnessed for measuring winding numbers in a polaritonic honeycomb lattice. Experimentally, this is realized in an hexagonal lattice of coupled micropillars obtained by etching a semiconductor planar microcavity confining exciton-polaritons (see Fig.~\ref{grapheneMCD} (a) and Supplemental materials). The coupling of the ground state of each pillar gives rise to two bands emulating the $\pi$ and $\pi^{*}$ bands of graphene: a momentum-resolved image of the emission is presented in Fig.~\ref{grapheneMCD} (b) showing the presence of Dirac cones. Throughout this work, we consider polaritonic states in the low power linear regime.
		
	Our technique for extracting the mean chiral displacement as a function of $k_{x}$ relies on photoluminescence (PL) measurements with combined real- and momentum-space resolution. The excitation is provided by a non-resonant CW laser focused on a single pillar that generates an incoherent wave-packet spanning both energy bands. We then measure its time-integrated emission profile along $y$, while filtering a well-defined momentum component along $x$. This is done using the optical imaging technique described in Fig.~\ref{grapheneMCD} (c), where a cylindrical lens (CL) with a curvature along $x$ is positioned at a focal distance of the Fourier plane (red dashed line) of collection lens L$_{1}$. The CL alters light trajectory only along $x$, providing a Fourier transform of the emission in this direction (upper panel); along $y$, L$_{1}$ and L$_{2}$ simply provide real space imaging of the emission (lower panel). Using a vertical slit at the imaging plane, this optical setup thus allows selecting a well-defined value of $k_{x}$, while simultaneously accessing the spatial profile along $y$.
		
	\begin{figure}[b]
		\centering
		\includegraphics[trim=0cm 0cm 0cm 0cm, width=0.9\columnwidth]{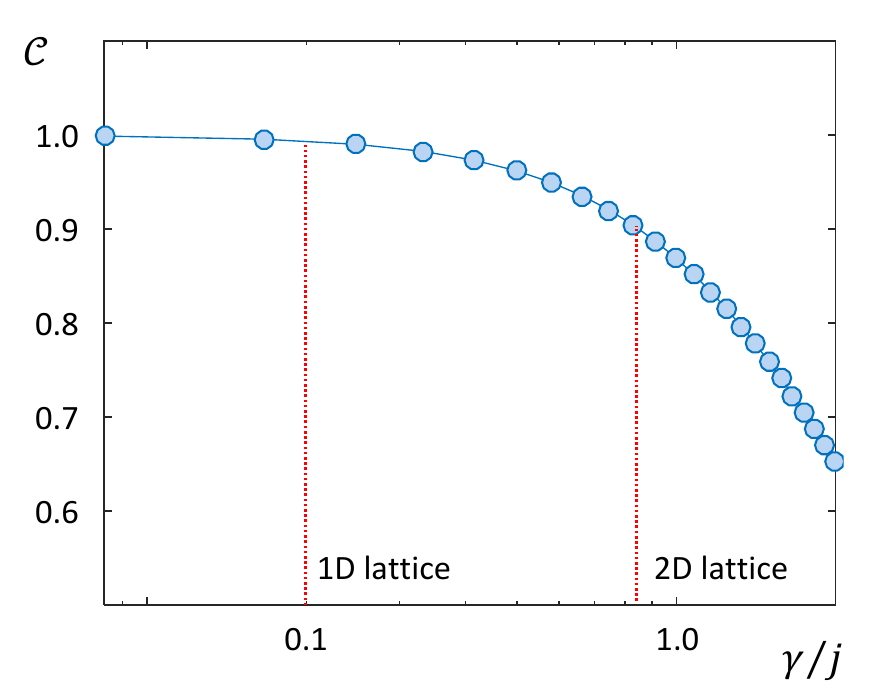}
		\caption{\textbf{Non-hermitian mean chiral displacement.} Calculated mean chiral displacement as a function of losses ($\gamma / j$) for a bearded edge in graphene, at $k_{x}=0$. The two red dashed lines indicated the typical values of losses for 1D and 2D polaritonic lattices.}
		\label{MCD_losses}
	\end{figure}
		
	\begin{figure*}
		\centering
		\includegraphics[trim=0cm 0cm 0cm 0cm,  width=0.9\textwidth]{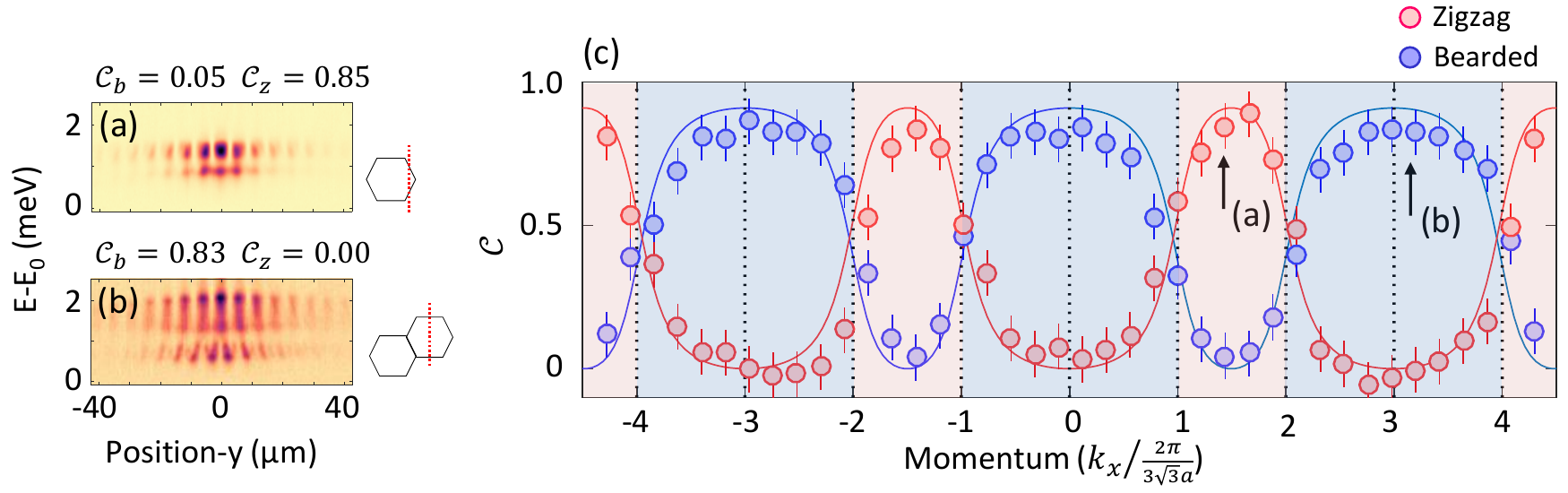}
		\caption{\textbf{Topological phase transitions in graphene.} (a)-(b) Spatially resolved (along $y$) emission spectra for two distinct momentum components $k_{x}$ schematically depicted in the inset. Values of $\mathcal{C}$ are provided above each panel. $E_{0}=1.571~\mathrm{eV}$. (c) Evolution of the mean chiral displacement as a function of $k_{x}$. Blue and red circles correspond to values associated to the two possible definitions of the unit cell: respectively bearded and zigzag. Blue (red) shaded areas correspond to regions where $\mathcal{W_{\mathrm{HC}}}=1$ for a unit cell definition compatible with zigzag (bearded) terminations. Dotted lines indicate the positions of Dirac cones. The solid line represents a theoretical calculation of the mean chiral displacement including losses. For details on the error analysis, see Supplementals materials.}
		\label{phaseTransitions}
	\end{figure*}
	
	Using this setup, Fig.~\ref{grapheneMCD} (d) presents a spatially-resolved (along $y$) emission profile for a position of the CL selecting momentum component $k_{x}\sim0$. This steady-state emission profile clearly describes the physics of a dimer chain with a lower bonding band and an upper anti-bonding band (the position of each effective site is depicted above the panel with the pumped site in red). From this intensity profile, it is then possible to extract the winding number by computing the mean chiral displacement\cite{Cardano2017, Maffei2018} (see Supplemental material):
	
	\begin{equation}
	\mathcal{C}(k_{x})=\int\limits dy ~ 2\Gamma (y)Y(y)I^{(\mathrm{int})}(y,k_{x})
	\label{mcd}
	\end{equation}
	
	\noindent where the integral is taken over the entire emitting region. $I^{(\mathrm{int})}(y,k_{x})$ is the normalized energy-integrated spatial profile of the emission (Fig.~\ref{grapheneMCD} (e)), $\Gamma(y)$ labels the sub-lattice index of the site (i.e. +1 (-1) for the A (B) sub-lattice) and $Y(y)$ labels the index of the unit cell (the wave-packet is created in the $0^{\mathrm{th}}$ unit cell). The product $\Gamma(y)Y(y)$ is a function describing the observable quantity associated to the chiral displacement operator $\hat{\Gamma}\hat{Y}$ (see Supplemental materials). 
	
	For a finite lattice, the unit cell is defined by the edge. However, for an infinite lattice, the two definitions shown in Fig.~\ref{sshGraphene} (a) are equally valid, and, in the same experiment, we can compute $\mathcal{W_{\mathrm{HC}}}$ for unit cells compatible with zigzag and bearded edges just by shifting the definition of $\Gamma$. Values of $\Gamma(y)Y(y)$ are presented in Fig.~\ref{grapheneMCD} (f) where the blue and red curves are compatible, respectively, with bearded and zigzag terminations. Computation of the mean chiral displacement, for $k_{x}=0$, leads to $\mathcal{C}_{z}=0.12(8)$ for the zigzag unit cell and $\mathcal{C}_{b}=0.84(8)$ for the bearded one, which clearly allows discriminating the two distinct topological phases.
	
	The observed deviation from a quantized winding number (i.e. either 0 or 1) is a direct consequence of the non-hermitian nature of exciton-polaritons: their finite lifetime prevents the wave-packet from reaching a fully balanced distribution over the two sub-lattices. Therefore, the emission profile does not exactly reflect the chiral symmetry of the underlying array. A derivation of the mean chiral displacement in the presence of losses is provided in the Supplemental Materials. Figure~\ref{MCD_losses} shows the evolution of this quantity, for bearded edges at $k_{x}=0$, as a function of the dissipation rate $\gamma$. It clearly shows a monotonous decay from $\mathcal{C}=1$ for a conservative system (i.e. $\gamma\rightarrow 0$), reaching a value of $\mathcal{C} = 0.91$ for typical dissipation rates of 2D polaritonic lattices, i.e. $\hbar\gamma\sim 150~\mu \mathrm{eV}$. For comparison purposes, measurement of the mean chiral displacement on a conventional 1D SSH lattice is provided in the Supplemental materials. For such 1D lattices, the polariton lifetime is typically an order of magnitude longer, thanks to the less aggressive etching technique required. As a result, $\mathcal{C}_{1}=0.05(8)$ and $\mathcal{C}_{2}=0.98(8)$ for the trivial and non-trivial definitions of the unit cell, in very good agreement with the expected winding numbers.
	
	It is then possible to access profiles associated to different $k_{x}$ values by laterally shifting the CL. Figure~\ref{phaseTransitions} (a) and (b) show spatially resolved emission spectra, at $k_{x}$ values indicated in Panel (c), which are qualitatively very different from the one measured at $k_{x}=0$. This reflects the fact that the effective coupling ratio ($J'/J$) changes with $k_{x}$. The emission pattern in Panel (a) exhibits almost flat bands, as one of the two coupling coefficients vanishes thus emulating the physics of a chain of uncoupled dimers. In Panel (b), $J'/J$ is negative, leading to a reversal of the bonding and anti-bonding bands.
	
	\begin{figure*}
		\centering
		\includegraphics[trim=0cm 0cm 0cm 0cm,  width=0.9\textwidth]{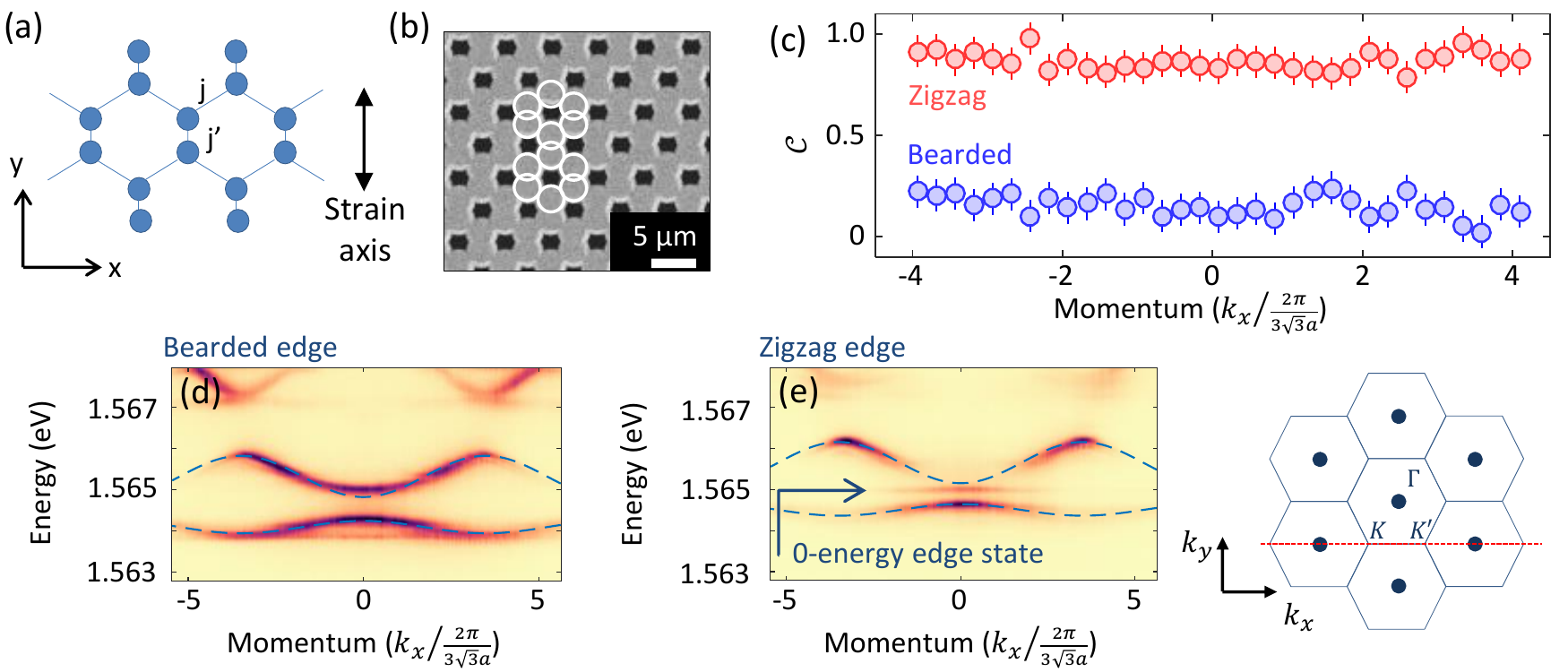}
		\caption{\textbf{Mean chiral displacement in a compressed honeycomb lattice.} (a) Schematic representation of the effect of compression on the lattice. The compression factor is defined as $\beta=j'/j$. (b) Top view SEM image of the compressed honeycomb lattice. White circles are added to indicate the positions of the pillars. (c) Evolution of the mean chiral displacement for both definitions of the unit cell. Blue circles correspond to the definition depicted in Fig.~\ref{sshGraphene} (a). (d)-(e) Momentum-resolved PL spectra measured at the boundary of the lattice, along a bearded (d) and a zigzag (e) termination. The cut in the BZ along which both images are taken is depicted in the right inset.}
		\label{grapheneStretched}
	\end{figure*}
		
	Figure~\ref{phaseTransitions} (c) present measured values of the mean chiral displacement as a function of $k_{x}$, obtained by scanning laterally the CL. Solid lines show numerical calculations of the mean chiral displacement including the effect of polariton lifetime (see Supplemental materials).  These measurements of the mean chiral displacement show good agreement with the predicted values of the winding number: each time $k_{x}$ crosses Dirac cones (indicated by vertical dotted lines), the values of $\mathcal{C}$ associated to each definition of the unit cell are exchanged, indicating a topological phase transition. Despite the presence of dissipation, these measured values of the mean chiral displacement present a clear contrast between high and low values, allowing to unambiguously identify the distinct topological phases. One important consequence of these transitions is the emergence (or disappearance) of 0-energy edge states as a function of $k_{x}$. Shaded blue (red) areas correspond to values of $k_{x}$ where zigzag (bearded) terminations exhibit edge states, and reported experimentally in Refs. \onlinecite{Plotnik2014, Milicevic2015}: these areas are indeed fully compatible with the measured values of $\mathcal{W_{\mathrm{HC}}}$.\\
	
	\noindent\textbf{Topological characterization of Dirac cones merging in compressed graphene}
	
	Having demonstrated the measurement of winding numbers in a honeycomb lattice using the mean chiral displacement, we now show its versatility by applying this technique to a lattice compressed beyond the merging transition of Dirac cones. It was shown, both theoretically\cite{Wunsch2008, Montambaux2009} and experimentally\cite{Tarruell2012, Rechtsman2013, Bellec2013, Bellec2014, Real2020}, that applying a uniaxial strain to honeycomb lattices along the $y$-axis by increasing the hopping amplitude $j'$ with respect to $j$, as depicted in Fig.~\ref{grapheneStretched} (a), shifts the position of the Dirac cones in the band structure along $k_{x}$. For strain coefficients $\beta=j'/j>1$ (i.e. compression), $K$ and $K'$ Dirac cones move toward each other, and merge beyond the critical value of $\beta=2$, leading to the opening of an energy gap for every $k_{x}$. This disappearance of gap closings eliminates the previously observed topological transitions between regions separated by Dirac cones.	
	
	To demonstrate this disappearance of phase transitions in critically compressed honeycomb lattices, we fabricate a lattice of coupled micropillars with center-to-center distances $d=2.4~\mu m$ and $d'=1.7~\mu m$ corresponding to $\beta=3$ (a SEM image of the lattice is shown in Fig.~\ref{grapheneStretched} (b)). Using the same measurement protocol as the one used for regular graphene, Fig.~\ref{grapheneStretched} (f) presents the evolution of the values of mean chiral displacement as a function of $k_{x}$, for unit cells compatible with zigzag and bearded terminations. Both values of the mean chiral displacement ($\mathcal{C}_{b}\sim0.1$ and $\mathcal{C}_{z}\sim0.9$, respectively) are now independent of $k_{x}$, indicating the disappearance of topological phase transitions as observed in Fig.~\ref{grapheneMCD} (f).
	
	As a result, bearded terminations never present edge states, and zigzag ones do for all values of $k_{x}$. This is confirmed by probing the emission near the boundaries of the lattice: Figs.~\ref{grapheneStretched} (d) and (e) present PL spectra as a function of momentum along bearded and zigzag terminations, respectively. The cut in momentum space along which these spectra are taken is depicted by the red line in the right inset. For the zigzag termination, an edge state band emerges in the centre of the gap, whereas no edge state is observed for the bearded termination. This provides a direct evidence of the bulk-edge correspondence in this lattice.\\
	
	\noindent\textbf{Conclusion and outlook}
	
	We demonstrated in this work a powerful approach for measuring 1D topological invariants from the bulk of 2D lattices presenting a chiral symmetry. These invariants are intimately related to the existence of edge states in honeycomb lattices. We have also studied critically compressed graphene, where topological phase transitions disappear when Dirac cones merge. One important next step is to extend our scheme to the study of other 2D chiral lattices, e.g. with flat bands\cite{Jiang2019, Lim2019}, involving higher-energy orbitals\cite{Milicevic2017, Milicevic2019a} or synthetic dimensions\cite{Ozawa2019a}. Furthermore, it would be interesting to explore how this scheme could be extended to the extraction of higher-order topological invariants\cite{Peterson2018, Serra-Garcia2018, Mittal2019, ElHassan2019} in chiral-symmetric lattices. Finally, the simplicity and versatility of our scheme opens the door to exploring more complex topological effects in polaritonic lattices, e.g. in the presence of disorder or interactions.\\
	
	\clearpage
	
	\noindent\textbf{Acknowledgements.}
	
	\noindent The authors acknowledge G. Montambaux for fruitful discussions. This work was supported by the H2020-FETFLAG project PhoQus (820392), the QUANTERA project Interpol (ANR-QUAN-0003-05), the French National Research Agency project Quantum Fluids of Light (ANR-16-CE30-0021), the French government through the Programme Investissement d'Avenir (I-SITE ULNE / ANR-16-IDEX-0004 ULNE) managed by the Agence Nationale de la Recherche, the French RENATECH network, the Labex CEMPI (ANR-11-LABX-0007), the CPER Photonics for Society P4S and the M\'etropole Europ\'eenne de Lille (MEL) via the project TFlight. A.D. and P.M. acknowledge financial support from the Spanish Ministry MINECO (National Plan 15 Grant: FISICATEAMO No.~FIS2016-79508-P and FIS2017-84114-C2-1-P, SEVERO OCHOA No.~SEV-2015-0522, FPI), European Social Fund, Fundaci\'o Cellex, Generalitat de Catalunya (AGAUR Grant No.~2017 SGR1341 and CERCA Program), ERC AdG NOQIA, the National Science Centre, Poland-Symfonia Grant No.~2016/20/W/ST4/00314, the ``Juan de la Cierva" fellowship (IJCI-2017-33180), ``Ram\'on y Cajal" program and EU FEDER Quantumcat. P.S.-J. acknowledges financial support from the Marie Sklodowska-Curie individual fellowship ToPol.\\
	
	\clearpage
	\onecolumngrid
	
	\setcounter{equation}{0}
	\setcounter{figure}{0}
	\setcounter{table}{0}
	\setcounter{section}{0}
	\renewcommand{\theequation}{S.\arabic{equation}}
	\renewcommand{\thefigure}{S\arabic{figure}}
	\renewcommand{\thetable}{S\arabic{table}}
	
	
	\begin{center}
		\large{\textbf{\textsc{Supplementary material: Measuring topological invariants in polaritonic graphene}}}
	\end{center}
	
	\section{Sample and experimental details}
	
	\subsection{Sample description and experimental technique}
	
	The lattices are etched out of a planar semiconductor cavity with high quality factor ($Q\sim 70 000$) consisting of a $\mathrm{Ga_{0.05}Al_{0.95}As}$ $\lambda /2$ layer embedded between two $\mathrm{Ga_{0.05}Al_{0.95}As/Ga_{0.2}Al_{0.8}As}$ Bragg mirrors formed from 28 (40) pairs in the top (bottom) mirror. Three sets of four GaAs quantum wells of 7~nm width are grown at the three central maxima of the electromagnetic field in the cavity, resulting in strong photon-exciton coupling exhibiting a 15~meV Rabi splitting. After the epitaxy, the cavity is processed by electron beam lithography and dry etching to form 2-dimensional lattices of overlapping cylindrical micropillars. For the regular honeycomb lattice, the diameter of the pillars is 2.7~$\mu\mathrm{m}$ and centre-to-centre distance is 2.4~$\mu\mathrm{m}$, allowing for the hopping of polaritons. For the compressed honeycomb lattice, the normal ($d$) and compressed ($d'$) centre-to-centre distances are respectively 2.4~$\mu\mathrm{m}$ and 1.7~$\mu\mathrm{m}$.
	
	Non-resonant PL measurements are realized with a single-mode CW laser at 745~nm. The emission is collected through a microscope objective and imaged on the entrance slit of a spectrometer coupled to a CCD camera with a spectral resolution of $\mathrm{\sim30~\mu eV}$, using the experimental setup depicted in Fig.~2 (c) of the main text. The sample is cooled down at $\mathrm{T=4~K}$.
	
	\subsection{Error analysis} 
	
	Error bars shown in Figs.~4 and 5 of the main text are obtained by evaluating the standard deviation of the extracted mean chiral displacement over several repetitions of the same measurements. They are of the order of 0.08, and do not change significantly as a function of $k_{x}$. 
	
\begin{figure*}[h]
		\includegraphics[width=0.9\linewidth]{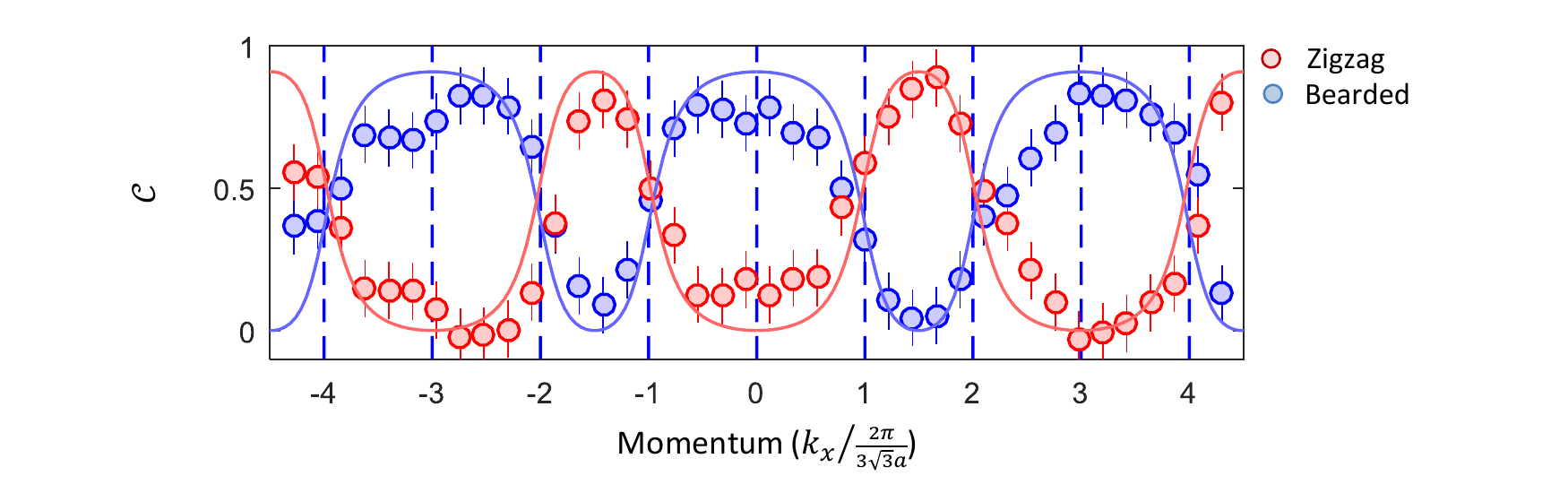}
		\caption{Evolution of the mean chiral displacement as a function of $k_{x}$ for regular graphene. Identical to Fig.~4 (c) of the main text, but without using a correction on the definition of the unit cell for each value of $k_{x}$, i.e. the unit cell is identically defined for each point.}
		\label{fig:uncorrected}
	\end{figure*}

	On top of this statistical error, several points were subjected to a systematic error. This second type of error was associated to a slight vertical shift of the image on the CCD camera when laterally moving the cylindrical lens. This shift, typically on the order of few pixels (corresponding to $\sim 100~\mathrm{nm}$), is most likely due to a slight misalignment in the imaging setup. In order to correct this effect when computing the mean chiral displacement, we have finely adjusted, for each $k_{x}$ point, the calibration relating the pixels of the camera to the imaged spatial position on the sample. This adjustment is necessary for properly defining the borders of each unit cell in Eq.~4 of the main text. Each calibration was done using spatially resolved photoluminescence (PL) spectra. Unadjusted data from Fig.~4 (c) of the main text, i.e. where the calibration is identical for each value of $k_{x}$, are presented in Fig.~\ref{fig:uncorrected}. We indeed see small systematic deviations of few clusters of points from the theoretical curve.
	
	\section{Time-integrated mean chiral displacement}
	
	The definition of the mean chiral displacement used in Refs. \onlinecite{Cardano2017, Maffei2018} considers the \textit{instantaneous} spatial profile of the Bloch modes in a 1D chiral lattice generated by a pulsed excitation:
	
	\begin{equation}
	\mathcal{C}(t)=\sum_{i}\bra{\psi_{i}(t)}2\hat{\Gamma_{i}} \hat{Y_{i}}\ket{\psi_{i}(t)}
	\end{equation}
	
	\noindent where $\ket{\psi_{i}(t)}$ corresponds to the instantaneous amplitude of the wave-function at the position of the $i^{th}$ site, and $\hat{\Gamma_{i}}$ and $\hat{Y_{i}}$ refer to the sub-lattice and unit cell operators. These previous works showed that, in the long time limit (i.e. for $t\gg \hbar \langle j \rangle ^{-1}$, where $\langle j \rangle $ is the average hopping amplitude of the SSH lattice), the instantaneous value of the mean chiral displacement converges toward the winding number. In this work, we consider time-integrated profiles of the Bloch modes under a CW excitation. We show hereafter that such steady-state values of $\bar{\mathcal{C}}$ as well lead to the winding number. 
	
	We consider the time-integrated mean chiral displacement :
	
	\begin{equation}
	\bar{\mathcal{C}}=\frac{1}{T}\int_0^T \mathcal{C}(t)dt=\int\limits dy ~ 2\Gamma (y)Y(y)I^{(\mathrm{int})}(y,k_{x})
	\end{equation}
	
	\noindent where $I^{(\mathrm{int})}(y,k_{x})$ is the normalized time-integrated intensity for each point $y$ and $k_{x}$. $\Gamma (y)$ and $Y(y)$ are functions corresponding to the classical counterpart of $\Gamma_{i}$ and $Y_{i}$ operators.
	
	For a 1D SSH lattice, this quantity can be readily computed using the following expression of the instantaneous mean chiral displacement derived in Ref.~\onlinecite{Cardano2017}:
	
	\begin{equation}
	\mathcal{C}(t)=\mathcal{W}-\int_{-\pi}^{\pi}\frac{dk}{2\pi} \cos(2tE(k))(\mathbf{n}(k)\times \partial_{k}{\mathbf{n}(k)})_z,
	\end{equation}
	
	\noindent where $\mathcal{W}$ is the winding number, $E(k)$ is the energy dispersion, and $\mathbf{n}(k)$ is defined as:
	
	\begin{equation}
	H_{\mathrm{SSH}}(k)=E(k)\mathbf{n}(k)\cdot\boldsymbol{\sigma}
	\end{equation}
	
	\noindent with $\boldsymbol{\sigma}=(\sigma_{x},\sigma_{y},\sigma_{z})$ is the Pauli vector and $H_{\mathrm{SSH}}$ the 1D SSH Hamiltonian. 
	
	We find the following expression for $\bar{\mathcal{C}}$
	
	\begin{equation}
	\begin{split}
	\bar{\mathcal{C}} &=\mathcal{W}-\frac{1}{2\pi T}\int_{-\pi}^{\pi}dk\,(\mathbf{n}(k)\times \partial_k{\mathbf{n}(k)})_z\int_0^Tdt \cos(2tE(k))\\
	&=\mathcal{W}-\frac{1}{2\pi T}\int_{-\pi}^{\pi}dk\,\frac{\sin(2 T E(k))}{2E(k)}  \, (\mathbf{n}(k)\times \partial_k{\mathbf{n}(k)})_z. 
	\end{split}
	\label{eq:integretad_mcd}
	\end{equation}
	
	In the limit of long integration times ($T\rightarrow\infty$), the second term in Eq. \eqref{eq:integretad_mcd} rapidly vanishes and $\bar{\mathcal{C}}$ converges to the winding number $\mathcal{W}$. In the main text, $\bar{\mathcal{C}}$ is simply referred to as $\mathcal{C}$.
	
	\section{Time-integrated mean chiral displacement in a 1D SSH lattice}
	
\begin{figure*}
		\includegraphics[width=0.9\linewidth]{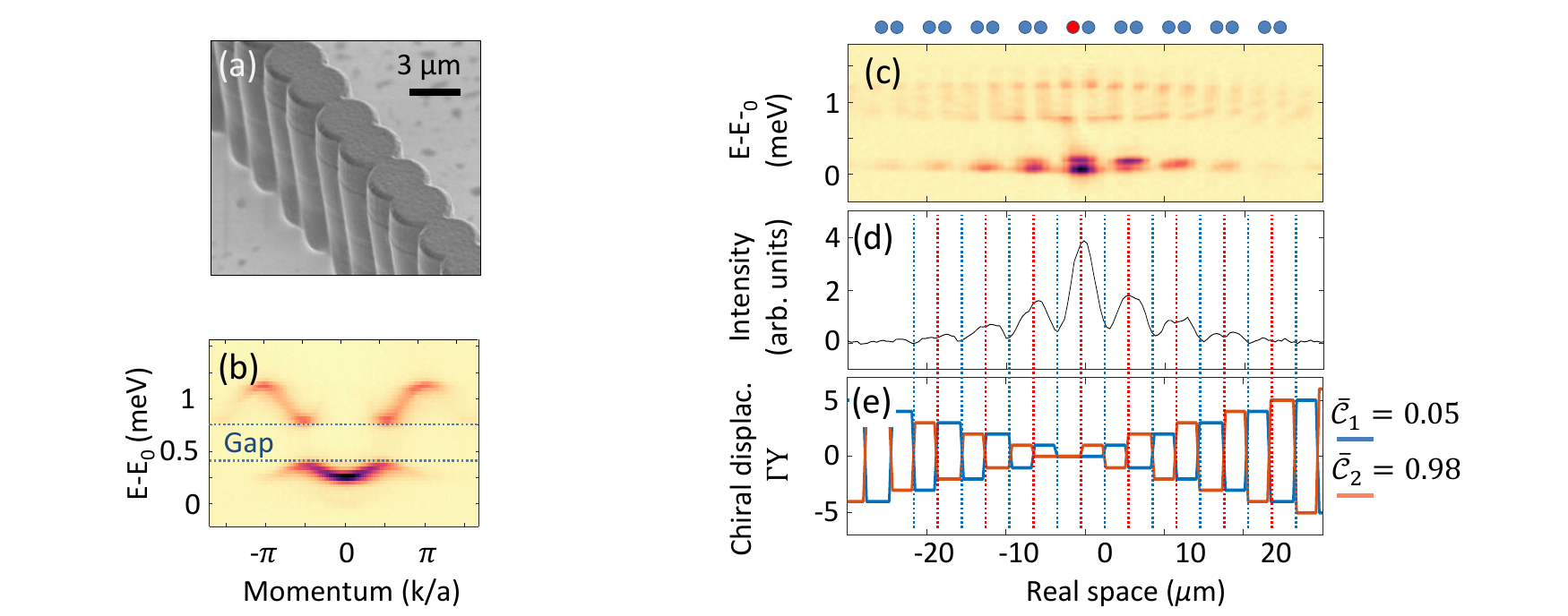}
		\caption{(a) SEM image of a 1D lattice of coupled micropillars emulating the SSH Hamiltonian. (b) Momentum-resolved PL spectra. (c) Spatially resolved emission spectra under a non-resonant CW excitation localized on a single pillar. The position of the pump is indicated in red above the panel. (d) Energy-integrated emission profile. (e) Spatial evolution of the chiral displacement $\Gamma Y$. Blue and red curves are defined for a unit cell presenting a strong and weak intra-cell coupling, respectively.}
		\label{fig:sshReal}
	\end{figure*}

	In order to experimentally benchmark this time-integrated version of the mean chiral displacement, we use a simple 1D lattice of coupled micropillars emulating the SSH Hamiltonian. This lattice is etched in a similar fashion to the honeycomb lattice in the main text. The pillars have a diameter of $\mathrm{3~\mu m}$ and are separated by a distance alternating between $\mathrm{2.2~\mu m}$ and $\mathrm{2.8~\mu m}$, in order to emulate the staggered hopping energies of the SSH Hamiltonian. A SEM image of this lattice is provided in Fig.~\ref{fig:sshReal} (a), and Fig.~\ref{fig:sshReal} (b) shows momentum-resolved emission spectra that clearly exhibit the two bands of the SSH model.
	
	To extract the integrated mean chiral displacement in this lattice, we use a non-resonant CW excitation localized on a single pillar, similarly as described in the main text. Fig.~\ref{fig:sshReal} (c) shows an image of time-integrated emission spectra as a function of spatial position, which corresponds to the steady-state emission profile of the system. Above the panel, a schematic representation of the position of the different pillars is added with the pumped pillar in red. Fig.~\ref{fig:sshReal} (d) shows the emission integrated in energy over the two bands, and Fig.~\ref{fig:sshReal} (e) the spatial evolution of the function $\Gamma(y) Y(y)$ for both definitions of the unit cell: red corresponds to the topological phase (weak intra-cell coupling) and blue corresponds to the trivial phase (strong intra-cell coupling). This emission profile yields mean chiral displacement values of 0.98 and 0.05, thus demonstrating a rapid convergence toward the winding number values of 1 and 0.
	
	This experimentally demonstrates the soundness of using the integrated mean chiral displacement for extracting the winding number of a chiral lattice.
	
	\section{Time-integrated mean chiral displacement with losses}
	
	As pointed out in the manuscript, one critical aspect of 2D polaritonic lattices, with respect to 1D lattices such as the one considered above, is that the lifetime of polaritons becomes non-negligible compared to the hopping characteristic time. The difference in lifetimes comes from the etching technique of 2D lattices that is more aggressive than for 1D lattices, thus generating a higher density of non-radiative recombination centres at the surface of pillars. 
	
	This decreased lifetime leads to the fact that the mean chiral displacement does not converge to the exact values of the winding number, because polaritons do not have time to reach a perfectly balanced population over the two sub-lattices. Therefore, the emission profile does not reflect perfectly the chiral symmetry of the underlying array. Hereafter, we provide a numerical analysis showing how the extracted value of the mean chiral displacement depends on losses. 
	
	We simulate the decay of polaritons out of the cavity in the form of photons by considering a wave-function whose amplitude decays exponentially, following $e^{-\gamma t/2}$ where $\gamma$ is the polariton lifetime In this case, the integrated mean chiral displacement is:
	
	\begin{equation}
	\begin{split}
	\bar{\mathcal{C}}&=\frac{1}{\int_0^T dt \braket{\psi}{\psi}}\int_0^T dt \bra{\psi} 2 \Gamma Y \ket{\psi}\\
	&=\frac{1}{\int_0^T dt e^{-\gamma t}}\int_0^T dt \int_{-\pi}^{\pi}\frac{dk}{2\pi} e^{-\gamma t}(1-\cos(2t E(k)))(\mathbf{n}(k)\times \partial_k{\mathbf{n}(k)})_z\\
	&=\mathcal{W}-\frac{1}{\int_0^T dt e^{-\gamma t}} \int_{-\pi}^{\pi}\frac{dk}{2\pi}(\mathbf{n}(k)\times \partial_k{\mathbf{n}}(k))_z  \int_0^T e^{-\gamma t} \cos(2t E(k))\\
	&=\mathcal{W}-\frac{\gamma}{1-e^{-\gamma T}}\int_{-\pi}^{\pi}\frac{dk}{2\pi} e^{-\gamma T} \, \frac{\gamma e^{\gamma T}-\gamma \cos(2 E(k) T)+2 E(k) \sin (2 E(k) T)}{\gamma^2+4 E(k)^2} (\mathbf{n}(k)\times \partial_k{\mathbf{n}(k)})_z. 
	\end{split}
	\label{eq:mcd_losses}
	\end{equation}
	
	This formula has an asymptotic limit for $T\rightarrow \infty$:
	
	\begin{equation}
	\lim_{T\rightarrow \infty} \bar{\mathcal{C}} = \mathcal{W}-\frac{1}{2\pi}\int_{-\pi}^{\pi}dk \frac{\gamma^2}{\gamma^2+4 E(k)^2}(\mathbf{n}(k)\times \partial_k{\mathbf{n}(k)})_z.
	\label{eq:mcd_losses_asympto}
	\end{equation}
	
	Figure~3 of the main text shows the computation of Eq. \eqref{eq:mcd_losses_asympto} as a function of losses ($\gamma/j$) for a 1D SSH Hamiltonian with a coupling ratio of $j/j'=0.5$, where $j$ and $j'$ respectively correspond to intra- and inter-cell coupling energies. This case also corresponds to $k_{x}=0$ in graphene. 
	
	\section{Effect of losses in polaritonic honeycomb lattices}
	
\begin{figure}[h]
		\includegraphics[width=0.9\linewidth]{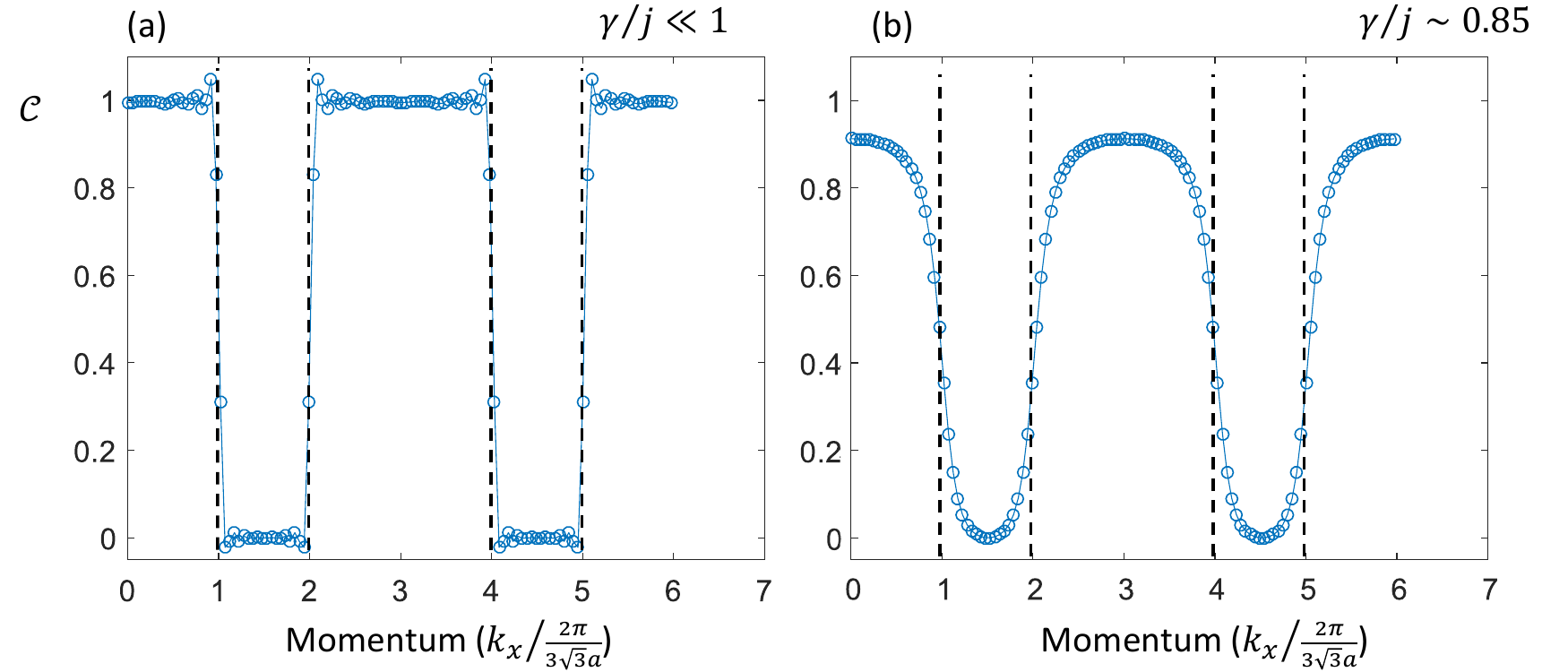}
		\caption{(a) Integrated mean chiral displacement (over a time $T=6\hbar/j$) in a honeycomb lattice as a function of momentum component $k_{x}$ for (a) negligible losses, and (b) losses compatible with a polaritonic honeycomb lattice ($t$ is the nearest neighbour coupling amplitude of the lattice).}
		\label{fig:integrated_mcd_losses}
	\end{figure}

	It is then possible to calculate the mean chiral displacement in a polariton honeycomb lattice (including losses) over the entire $k_{x}$ space by adapting Eq. \eqref{eq:mcd_losses_asympto} to:
	
	\begin{equation}
	\lim_{T\rightarrow \infty} \bar{\mathcal{C}}(k_{x}) = \mathcal{W}(k_{x})-\frac{1}{2\pi}\int_{-\pi}^{\pi}dk_{y} \frac{\gamma^2}{\gamma^2+4 E(\vec{k})^2}(\mathbf{n}(\vec{k})\times \partial_{k_{y}}{\mathbf{n}(\vec{k})})_z,
	\label{eq:mcd_losses_asympto_graph}
	\end{equation}
	
	\noindent with $E=E(\vec{k})$, the energy dispersion of graphene. Figures~\ref{fig:integrated_mcd_losses} (a) and (b) show the result of this equation for a ratio $\gamma/j \ll 1$ (i.e. negligible losses) and $\gamma/j=0.85$ which is compatible with experimentally extracted values for the specific polaritonic honeycomb lattice used in this work, i.e. $\gamma=150~\mu eV$ and $j=180~\mu eV$. 
	
	The consequences of losses are twofold. First, the mean chiral displacements never reaches unity, but a maximum value of $\bar{\mathcal{C}}=0.91$. Secondly, the evolution of the mean chiral displacement is less abrupt when crossing the Dirac cones (indicated by dashed lines). Both of these effects are in good agreement with measured values presented in the main text: the curve in Fig.~\ref{fig:integrated_mcd_losses} (b) corresponds to the theoretical one provided in Fig.~4 (c) of the main text. 
	
	\clearpage
	\twocolumngrid
	
	
	
	%

\end{document}